\documentclass[12pt]{article}                  
\usepackage{osajnl}
\usepackage{overcite,hyperref}

\begin{document}
\def\baselinestretch{1.2}

\title{High-precision Absolute Distance Measurement 
using Dual-Laser Frequency Scanned Interferometry Under
Realistic Conditions}

\author{Hai-Jun Yang$^*$ and Keith Riles$^\dagger$ \\
($^*$yhj@umich.edu, $^\dagger$kriles@umich.edu)}


\affiliation{Department of Physics, University of Michigan, Ann Arbor, MI 48109-1120, USA}

\begin{abstract}
In this paper, we report on new high-precision absolute distance 
measurements performed with frequency scanned interferometry
using a pair of single-mode optical fibers. Absolute distances were determined
by counting the interference fringes produced while scanning the frequencies
of the two chopped lasers.
High-finesse Fabry-Perot interferometers were used to determine frequency
changes during scanning. Dual lasers with oppositely scanning directions, 
combined with a multi-distance-measurement technique previously
reported, were used to cancel drift errors and 
to suppress vibration effects and interference fringe uncertainties.
Under realistic conditions, a precision about 0.2 microns was achieved for
a distance of 0.41 meters.
\end{abstract}


\maketitle 


\section{Introduction}

We reported previously on single-laser measurements of
absolute distance with a frequency scanned interferometry apparatus,
carried out under controlled laboratory conditions\cite{fsi05}. Here we report
on dual-laser measurements carried out under less favorable conditions,
more representative of the interior of a high energy physics detector
at a collider.
The motivation for these studies is to design a novel optical
system for quasi-real time alignment of tracker detector elements
used in High Energy Physics (HEP) experiments. A.F. Fox-Murphy {\em et.al.}
from Oxford University reported their design of a frequency
scanned interferometer (FSI) for precise alignment of the ATLAS
Inner Detector \cite{ox98,ox04,coe2001}. Given the demonstrated need for
improvements in detector performance, we plan to design and prototype an
enhanced FSI system to be used for the alignment of tracker
elements in the next generation of electron-positron Linear
Collider detectors\cite{ilc}. Current plans for future detectors
require a spatial resolution for signals from a tracker detector,
such as a silicon microstrip or silicon drift detector,
to be approximately 7-10 $\mu m$\cite{orangebook}. To achieve this
required spatial resolution, the measurement precision of absolute
distance changes of tracker elements in one dimension should be no worse than
about 1 $\mu m$. Simultaneous measurements from hundreds of interferometers
will be used to determine the 3-dimensional positions of the
tracker elements.

Detectors for HEP experiment must usually be operated remotely for 
safety reasons because of intensive radiation, high voltage or strong magnetic
fields. In addition, precise tracking elements are typically surrounded by
other detector components, making access difficult. For practical HEP application
of FSI, optical fibers for light delivery and return are therefore necessary.
The power of front-end electronics depends on the event occupancy levels and 
trigger rates during the collider and detector operation. It is possible that
the temperature distribution between heat sources and
cooling system will vary enough to cause significant shape changes of the
silicon detector in a relatively short period of time. Hence, it is critical to 
design a high precision optical alignment system that performs well under
unfavorable environmental conditions.

Absolute distance measurements using FSI under controlled conditions 
have been reported previously\cite{stone99,dai98,barwood98,bechstein98,thiel95,kikuta86}
by other research groups.

The University of Michigan group has constructed several demonstration
Frequency Scanned Interferometer (FSI) systems with the laser
light transported by air or single-mode optical fiber, using single-laser and dual-laser
scanning techniques for initial feasibility studies.
Absolute distance was determined by counting the interference
fringes produced while scanning the laser frequency.
The main goal of the demonstration systems has been to determine the
potential accuracy of absolute distance
measurements that could be achieved under both controlled and realistic conditions.
Secondary goals included estimating the effects of vibrations and studying
error sources crucial to the absolute distance accuracy.
Two multiple-distance-measurement analysis techniques
were developed to improve distance precision and to extract the amplitude
and frequency of vibrations.
Under well controlled laboratory conditions, 
a measurement precision of $\sim$ 50 nm was achieved for
absolute distances ranging from 0.1 meters to 0.7 meters by using the first
multiple-distance-measurement technique (slip measurement window with fixed size)\cite{fsi05}.
The second analysis technique (slip measurement window with fixed start point) 
has the capability to
measure vibration frequencies ranging from 0.1 Hz to 100 Hz with amplitude
as small as a few nanometers, without a {\em priori} knowledge\cite{fsi05}.
The multiple-distance-measurement analysis techniques are well suited for
reducing vibration effects and uncertainties from fringe \& frequency determination,
but do not handle well the drift errors such as from thermal effects.

The dual-laser scanning technique
was pioneered by the Oxford group for alignment of the ATLAS Semi-conductor tracker;
it was demonstrated that precisions of better than 0.4 $\mu$m and 0.25 $\mu$m for 
distances of 0.4 m and 1.195 m, respectively\cite{ox04}.
In our recent studies, we combine our multi-distance-measurement analysis technique
(slip measurement window with fixed size) reported previously
with the dual-laser scanning technique to improve the absolute distance measurement 
precision. The multi-distance-measurement technique is effective in reducing
uncertainties from vibration effects and interference fringe determination, while
the dual-laser scanning allows for cancellation of drift errors.
We report here on resulting absolute distance measurement precisions 
under more realistic conditions than in our previous, controlled-environment
measurements.

\section{Principles}
We begin with a brief summary of the principles of frequency scanned
interferometry and our single-laser measurement technique.
The intensity $I$ of any two-beam interferometer can be expressed as
$
I = I_1 + I_2 + 2\sqrt{I_1 I_2} \cos(\phi_1 - \phi_2)
$, where $I_1$ and $I_2$ are the intensities of the two combined beams,
and $\phi_1$ and $\phi_2$ are the phases.
Assuming the optical path lengths of the two beams are $D_1$ and
$D_2$, the phase difference is
$\Phi = \phi_1 - \phi_2 = 2\pi |D_1 - D_2|(\nu/c)$,
where $\nu$ is the optical frequency of the laser beam, and c is
the speed of light.

For a fixed path interferometer, as the frequency of the laser is
continuously scanned, the optical beams will
constructively and destructively interfere, causing ``fringes''.
The number of fringes $\Delta N$ is $\Delta N = D\Delta\nu/c$,
where $D$ is the optical path difference between the two beams,
and $\Delta\nu$ is the scanned frequency range. The optical path
difference (OPD for absolute distance between beamsplitter and retroreflector)
can be determined by counting interference fringes while scanning the laser frequency.

If small vibration and drift errors $\epsilon(t)$ occur during the laser scanning, then
$\Phi(t) =  2\pi \times [D_{true}+\epsilon(t)] \times \nu(t)/c$,
$\Delta N = [\Phi(t) - \Phi(t0)]/2\pi =
D_{true}\Delta\nu/c  + [\epsilon(t)\nu(t)/c - \epsilon(t0)\nu(t0)/c]$,
Assuming $\nu(t) \sim \nu(t0) = \nu$ and $\Delta \epsilon = \epsilon(t) - \epsilon(t0)$,
the measured distance can be written as,
$$
\begin{array}{c}
D_{measured} = \Delta N / (\Delta\nu/c) = D_{true} + \Delta\epsilon \times \Omega.
\end{array}
\eqno{(1)}
$$
where $\Omega$ is a magnification factor: $\Omega = \nu/\Delta\nu$.

\begin{figure}[htbp]
\begin{center}
\scalebox{0.6}{\includegraphics{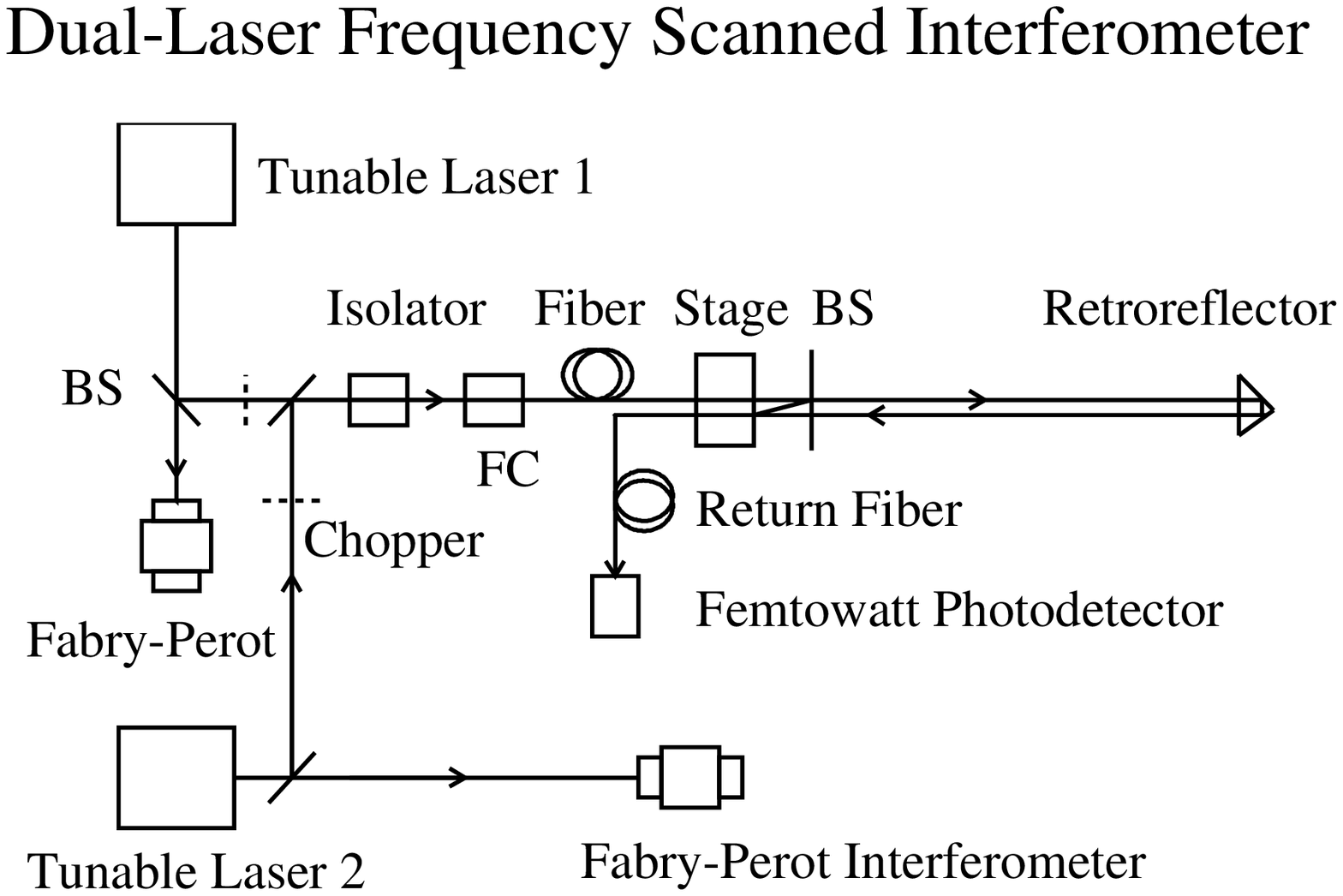}}
\scalebox{0.5}{\includegraphics{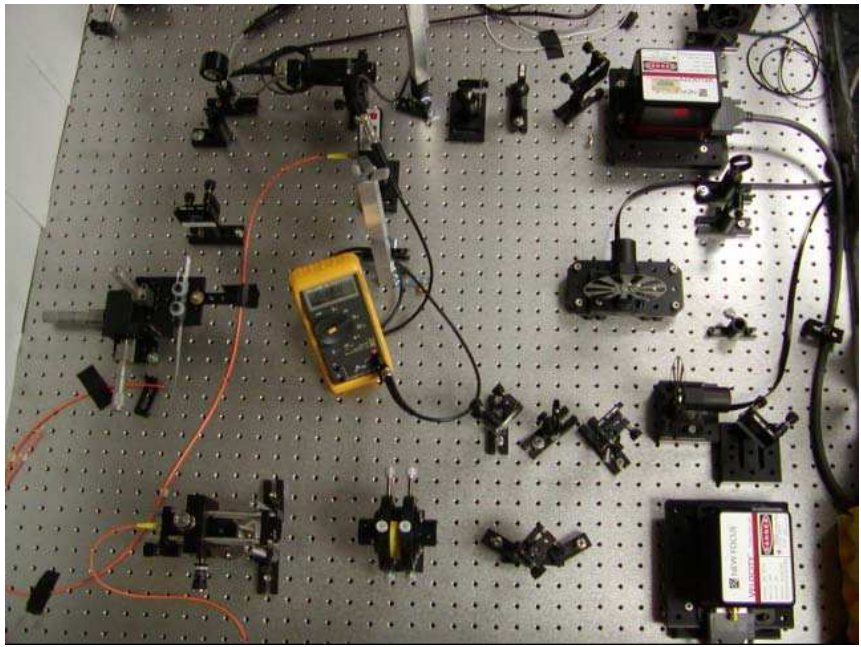}}
\caption{Schematic of the dual-laser FSI system.}
\end{center}
\end{figure}

\section{Demonstration System of FSI}

A schematic of the FSI system with a pair of optical fibers is shown in Figure 1.  
The light sources are two New Focus Velocity 6308 tunable lasers 
(Laser 1 - 665.1 nm $<\lambda<$ 675.2 nm; Laser 2 - 669.2 nm $<\lambda<$ 679.3 nm ). 
Two high-finesse ($>200$) Thorlabs SA200 Fabry-Perot are used to measure the frequency range 
scanned by the laser. The free 
spectral range (FSR) of two adjacent Fabry-Perot peaks is 1.5 GHz, which corresponds 
to 0.002 nm. A Faraday Isolator was used to reject light reflected back into
the lasing cavities. The laser beams were coupled into a single-mode optical fiber
with a fiber coupler. Data acquisition is based on a National Instruments DAQ
card capable of simultaneously sampling 4 channels at a rate of 5 MS/s/ch with
a precision of 12-bits.  Omega thermistors with a tolerance of 0.02 K and a
precision of 0.01 $mK$ are used to monitor temperature.  The apparatus
is supported on a damped Newport optical table.

The beam intensity coupled into the return optical fiber is very weak, requiring
ultra-sensitive photodetectors for detection. Given the low light
intensity and the need to split into many beams to serve a set
of interferometers, it is vital to increase the geometrical efficiency.
To this end, a collimator is built by placing an optical fiber in a ferrule 
(1mm diameter) and gluing one end of the optical fiber to a GRIN lens. 
The GRIN lens is a 0.25 pitch lens with  0.46 numerical aperture, 1 mm diameter 
and 2.58 mm length which is optimized for a wavelength of 630nm. 
The density of the outgoing beam from the optical 
fiber is increased by a factor of approximately 1000 by using a GRIN lens. 
The return beams are received by another optical fiber and amplified by a 
Si femtowatt photoreceiver with a gain of $2 \times 10^{10} V/A$.

\section{Dual-Laser Scanning Technique}

A dual-laser FSI system was built in order to reduce drift error and slow
fluctuations occurring during the laser scan, as shown in Figure.1.
Two lasers are operated simultaneously; the two laser beams are coupled into one optical
fiber but isolated by using two choppers.
The advantage of the dual-laser technique comes from cancellation of systematic
uncertainties, as indicated in the following.
For the first laser, the measured distance
$D_1 = D_{true} + \Omega_1 \times \Delta\epsilon_1$,
and $\Delta\epsilon$ is drift error during the laser scanning.
For the second laser, the measured distance
$D_2 = D_{true} + \Omega_2 \times \Delta\epsilon_2$.
Since the two laser beams travel the same optical path during the same period,
the drift errors $\Delta\epsilon_1$ and $\Delta\epsilon_2$ should be very comparable.
Under this assumption, the true distance can be extracted using the formula
$D_{true} =  (D_2-\rho \times D_1)/(1-\rho)$, where, $\rho = \Omega_2/\Omega_1$,
the ratio of magnification factors from two lasers. If two similar lasers
scan the same range in opposite directions simultaneously, then $\rho \simeq -1.0$, and $D_{true}$
can be written as,
$$
\begin{array}{c}
D_{true} =  (D_2-\rho \times D_1)/(1-\rho) \simeq (D_2 + D_1)/2.0
\end{array}
\eqno{(2)}
$$

Unfortunately, there are disadvantages too in the dual-laser technique. 
Because the laser beams are isolated by periodic choppers, only half the fringes are
recorded for each laser, degrading the distance measurement precision, as shown in Figure 2.
Missing fringes during chopped intervals for each laser must be
recovered through robust interpolation algorithms. Based on our studies, the number of
interference fringes in a time interval with fixed number of Fabry-Perot peaks is stable.
The measured number of fringes is nearly always within 0.5 (typically within 0.3)
of the expected fringe number, which enables us to estimate the number of fringes in
the chopper-off slots (when the laser beam is blocked by the chopper). In order to determine the
number of fringes in one chopper-off slot, we need to identify two Fabry-Perot peaks within
the two adjacent chopper-on slots closest to the chopper-off slot. If the fringe phases at
the two Fabry-Perot peaks positions are $I+\Delta I$ and $J + \Delta J$, where I and J
are integers, $\Delta I$ and $\Delta J$ are fraction of fringes; then the number
of true fringes can be determined by minimizing the quantity $|N_{correction} + (J+\Delta J) -
(I + \Delta I) - N_{expected-average}|$, where $N_{correction}$ is an integer
used to correct the fringe number in the chopper off slot, $N_{expected-average}$ is
the expected average number of fringes, based on a full laser
scanning sample shown in Figure 3. 

For the example shown in Figure 2, 
the number of Fabry-Perot intervals is 5 in the chopper-off slot, so the 
expected number of fringes shown in the 2nd plot of Figure 3 is about
20.6; The number of fringes measured in the chopper-off slot is 3.55,
$N_{correction}$ is 17 based on the above formula, meaning 
17 fringes are missed in the chopper-off slot. So the total number of
fringes in this slot is 20.55, well within the expected range $20.6 \pm 0.3$.

\begin{figure}[htbp]
\centerline{\scalebox{0.5}{\includegraphics{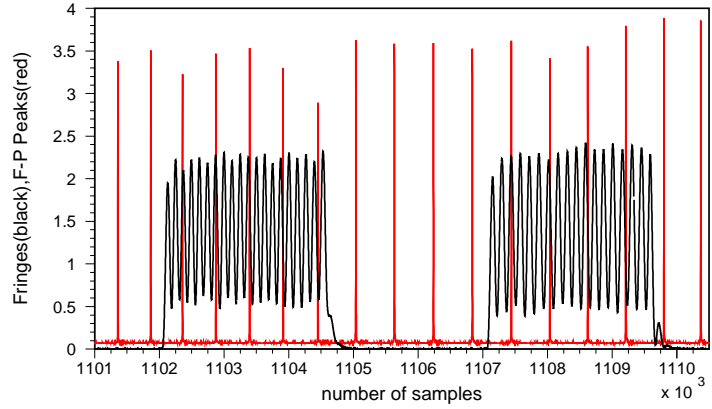}}}
\caption{Sample fringes and Fabry-Perot peaks from the first laser with chopping and with
the second laser turned off.}
\end{figure}

\begin{figure}[htbp]
\centerline{\scalebox{0.5}{\includegraphics{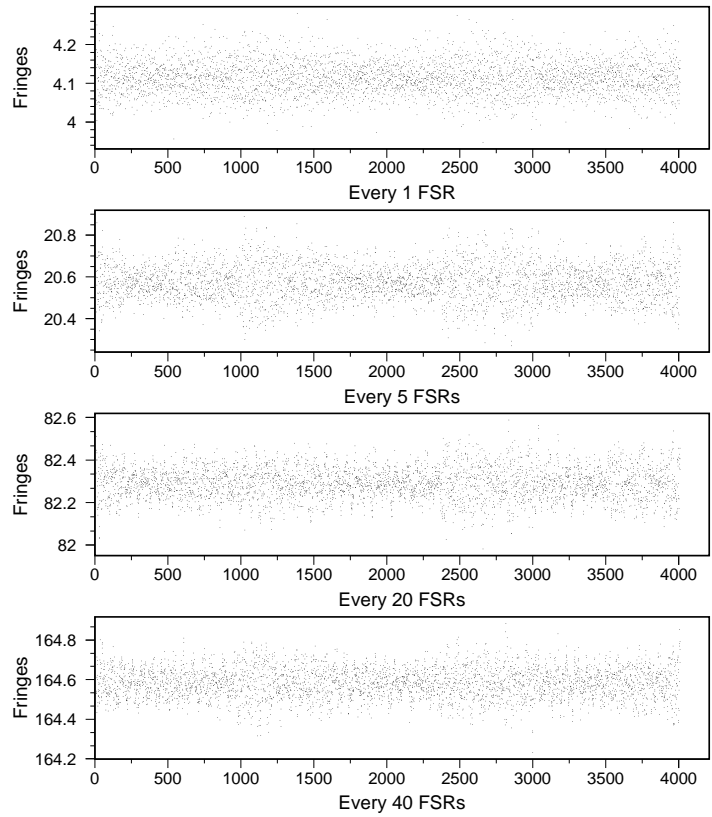}}}
\caption{The number of fringes for fixed numbers of Fabry-Perot intervals (1,5,20,40) from
full scan data.}
\end{figure}

\begin{figure}[htbp]
\centerline{\scalebox{0.5}{\includegraphics{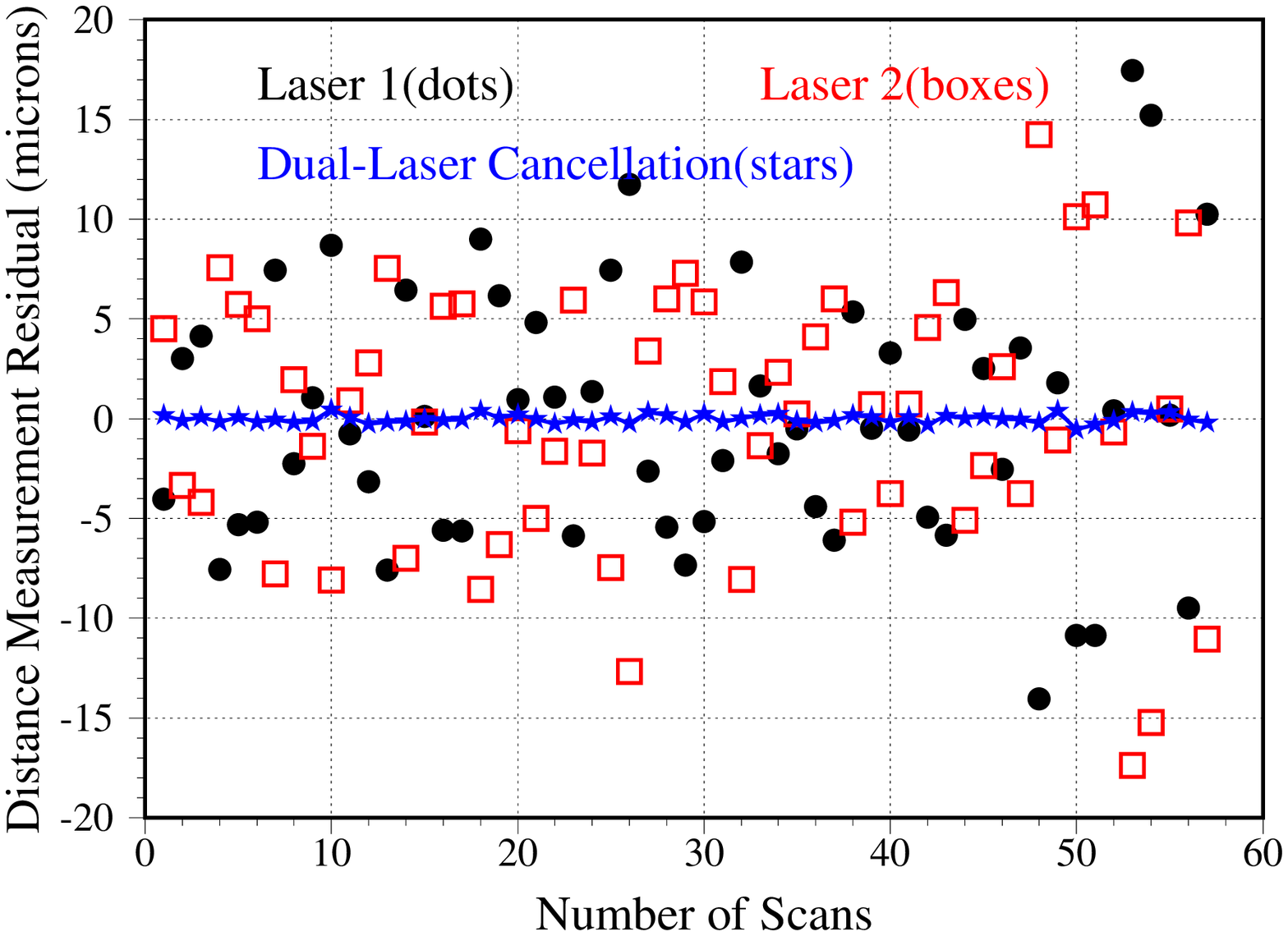}}}
\caption{Distance measurement residuals versus number of scans for dual-laser scanning data.
The measurement residuals refer to differences between measured distances and the arithemetic
averages of measured distances from each set of 10 scans under same conditions.
The number of distance measurements/scan is 2000 for these residuals.
Dots show distance measurement residuals from laser 1 and boxes from laser 2.
The distance measurement residuals with dual-laser cancellation are shown in stars.
}
\end{figure}

\section{Absolute Distance Measurement}

Under realistic and deliberately generated hostile conditions of 
air flow and vibration, 40 dual-laser-scan data samples were collected as listed in the following. 
The ``box'' refers to a nested enclosure of plexiglass box
covering the optical table and PVC pipe surrounding the interferometer beam path, used in
previous measurements\cite{fsi05} to isolate the interferometer from environmental fluctuations. 
Both the cover of the box and the pipe are removed for the measurements described here.

\begin{itemize}
\item with open box (10 scans) and ambient room environment, 

\item with open box and a chassis cooling fan (9 cm diameter) blowing air toward the beam splitter 
about 0.93 meters away from the retroreflector along the FSI (10 scans),

\item with open box and fan off (10 scans), 

\item with open box and a vibration source (10 scans), with a PI piezoelectric translator (P-842.10) 
used to generate controlled vibrations with frequency of 1 Hz and amplitude of about 0.15 microns.
(10 scans were collected, but 2 scans were found to have suffered a power glitch on one laser, 
invalidating the distance reconstructions. These 2 scans were excluded from the analysis.)

\end{itemize}

To verify correct tracking of large thermal drifts, another set of scans was carried out
after remounting the interferometer on a $1' \times 2' \times 0.5''$ Aluminum breadboard (Thorlabs)
and placing the breadboard on a heating pad (Homedics).
Twenty dual-laser-scan data samples with the heating pad off and on
were collected as listed in the following,

\begin{itemize}

\item with open box and the heating pad off (10 scans with one scan excluded because of a 
power glitch), 

\item  with open box and the heating pad on (10 scans). 

\end{itemize}

The two lasers were scanned oppositely with scanning
speeds of $\pm$0.4 nm/s and a full-scan time of 25 seconds.
The choppers have two blades with operation frequency of 20 Hz for
these dual-laser scans.
The measured precision is found to vary from about 3 to 11 microns if
we use the fringes of these data samples from only one laser
for a measured distance of 0.41 meters. The single-laser 
multi-distance-measurement technique
does not improve the distance measurement precision in open box data
because drift error dominates.
If we combine the measured distances from two lasers using Eq.(2), then 
the dual-laser measurement precision is found to be about 0.2 microns for a number of 
multi-distance-measurements larger than 500, as shown in Figure 4 and Table 1. 
From Table 1, it is apparent that if we use fewer distance measurements/scan 
or a single-distance-measurement, then the measured precision worsens significantly.
Combining the dual-laser scanning technique and multi-distance-measurement technique
ensures that vibrations, fringe uncertainties and drift errors are greatly
suppressed.

Under nearly identical conditions, the measured distances from laser 1 and laser 2 are slightly different,
as expected from their slightly different FSRs (taken to be 1.5 GHz for the distance measurements) of 
the two Fabry-Perot interferometers. Since a high precision wavemeter is not available in our 
laboratory currently, we cannot calibrate the FSRs of two Fabry-Perot interferometers precisely. 
Instead, we use the measured distances from two lasers to determine the ratio
(1.0005928) between the FSRs of the two Fabry-Perot interferometers, 
and then normalize all measured distances from the 2nd laser using the same factor. 
This correction method was validated previously with a different interferometer configuration by comparing
the normalization factor to that inferred from measuring the transmission peaks of the two 
Fabry-Perot interferometers  when monitoring the same laser. That check was limited in 
precision, however, to no better than 4 ppm.


In order to verify the correct tracking of large thermal drifts, we placed a heating pad 
on the Aluminum breadboard.
Thermistors are taped to the breadboard but electrically isolated.
The true temperature of the breadboard is inferred to be the 0.6 $\pm 0.1 ~^oC$ higher
than the measured temperature from extrapolation of temperature gradient from 3 external thermistors.
Four independent tests with different temperature increases are made, the results are shown in Table 2.
Each test has 20 dual-laser scans for the distance measurement before (10 scans) and after (10 scans) 
the temperature increase. 
The expected distance changes agree well with the measured distance changes due to thermal extension.
The uncertainty of the expected distance change comes mainly from the large error in temperature measurement
of the Aluminum breadboard.
Temperature fluctuations are found to be about $0.1 \sim 0.3 ~^oC$ in half a minute with the heating 
pad on (low - high level), about $ 2 - 6$ times larger than the temperature fluctuations with open box 
and the heating pad off.

In order to verify the distance measurement from FSI, a pizeoelectric transducer 
(P-842.10 from PI, with 20\% tolerance) was used to generate controlled 
position shifts of the retroreflector. 
For instance, for an input voltage of 13.13 volts, the expected distance change is
1.97 $\pm$ 0.39 microns. The measured distance change under controlled conditions 
is 2.23 $\pm$ 0.07 microns, in good agreement.

\section{Error Estimation}

Based on the error estimation we reported previously\cite{fsi05}, the
distance measurement statistical uncertainty is about 0.05 microns under 
well controlled, closed box conditions.
For the dual-laser scanning technique, only about 40\% of the independent fringe
measurements can be used in the multi-distance-measurement, giving an expected statistical uncertainty
of the distance measurement from each laser of no better than about $0.05 \times \sqrt{1./0.4} \sim 0.08$ microns.
The measurement errors from laser 1 and laser 2 propagate to the final distance
measurement using Eq.(2), leading to a corresponding error of about $0.08*\sqrt{2}/2 \sim 0.057$ microns.

Since manual scan starts give laser start times that may differ by as much as 0.5-1.0 seconds,
expected differences in drift errors are expected to differ by as much as 2\%-4\%.
Typically, the drift errors from single laser are found to be about 3 $\sim$ 11 microns 
for different samples under open-box conditions,
as shown in Table 1; so the expected uncertainty ranges from 0.03 $\sim$ 0.22 microns. 

Combining all above errors, the expected distance measurement statistical 
uncertainty ranges as high as 0.23 microns,
consistent with the measured variations.

Some other sources can contribute to systematic bias in the absolute distance measurement.
The major systematic bias comes from the uncertainty in the FSR of the Fabry-Perot used to 
determine the scanned frequency range. A high precision wavemeter(e.g. $\Delta \lambda / \lambda \sim 10^{-7}$) 
was not available for the measurements described here. The systematic bias from the 
multiple-distance-measurement technique is typically less than 50 nanometers.
The systematic bias from uncertainties in temperature, air humidity and barometric pressure scale are 
estimated to be negligible.

\begin{table}
\begin{tabular}{|c|c|c|c|c|c|c|c|c|c|} \hline
Data & Scans & Conditions & Distance(cm) & \multicolumn{6}{|c|}{Precision($\mu$m) for multi-dist.-meas./scan} \\ \cline{5-10}
     &       & open box   & from dual-laser &  2000 & 1500 & 1000 & 500  & 100  & 1     \\ \hline\hline
L1   & 10    & open box   & --           &  5.70 & 5.73 & 6.16 & 6.46 & 5.35 & 6.64  \\ \hline
L2   & 10    & open box   & --           &  5.73 & 5.81 & 6.29 & 6.61 & 5.66 & 6.92  \\ \hline
L1+L2& 10    & open box   & 41.13835     &  0.20 & 0.19 & 0.18 & 0.21 & 0.39 & 1.61  \\ \hline\hline

L1   & 10    & with fan on & --       &  5.70 & 4.91 & 3.94 & 3.49 & 3.29 & 3.04  \\ \hline
L2   & 10    & with fan on & --       &  5.70 & 5.19 & 4.23 & 3.78 & 3.21 & 6.07  \\ \hline
L1+L2& 10    & with fan on & 41.13841 &  0.19 & 0.17 & 0.20 & 0.22 & 0.31 & 3.18  \\ \hline\hline

L1   & 10    & with fan off & --          &  6.42 & 5.53 & 4.51 & 3.96 & 4.41 & 3.36  \\ \hline
L2   & 10    & with fan off & --          &  6.81 & 5.93 & 4.86 & 4.22 & 4.63 & 5.76  \\ \hline
L1+L2& 10    & with fan off & 41.13842    &  0.20 & 0.20 & 0.26 & 0.19 & 0.27 & 2.02  \\ \hline\hline

L1   & 8     & with vibration on$^*$
 & --      & 4.73 & 4.82 & 3.60 & 3.42 & 4.62 & 8.30 \\ \hline
L2   & 8     & with vibration on & --      & 4.72 & 4.66 & 3.66 & 3.65 & 4.63 & 5.56 \\ \hline
L1+L2& 8     & with vibration on & 41.09524 & 0.17 & 0.21 & 0.17 & 0.15 & 0.39 & 1.75  \\ \hline\hline

L1   & 9     & with heating pad off$^\dagger$ & --  & 3.88 & 3.90 & 3.57 & 3.65 & 3.28 & 3.84 \\ \hline
L2   & 9     & with heating pad off & --  & 4.01 & 4.01 & 3.64 & 3.55 & 3.25 & 4.66 \\ \hline
L1+L2& 9     & with heating pad off & 40.985122 & 0.14 & 0.14 & 0.11 & 0.12 & 0.19 & 1.86 \\ \hline\hline

L1   & 10    & with heating pad on & -- & 11.39 & 11.15 & 10.05 & 7.44 & 6.24 & 5.04 \\ \hline
L2   & 10    & with heating pad on & -- & 11.42 & 11.21 & 10.23 & 7.39 & 6.47 & 6.30 \\ \hline
L1+L2& 10    & with heating pad on & 40.987189 & 0.32  & 0.19 & 0.20 & 0.19 & 0.20 & 1.24 \\ \hline

\end{tabular}
\caption{Distance measurement precisions using the multiple-distance-measurement
and dual-laser scanning techniques. All 57 scans are for dual-laser scanning data.
Rows starting with L1 or L2 show results using fringes and Fabry-Perot information from
the 1st or 2nd laser only to make distance measurement, L1+L2 shows results by combining
measurement distances from both lasers to cancel the drift errors. 
{\it {
$^{(*)}$ Attaching the piezoelectric transducer required disturbing the position of the retroreflector.
$^{(\dagger)}$ The last two sets of data were taken four months after the
first four data sets, the beamsplitter and retroreflector are moved from optical table to the Aluminum
breadboard.}}
}
\end{table}

\begin{table}
\begin{center}
\begin{tabular}{|c|c|c|} \hline
$\Delta T  (^oC)$ & $\Delta R_{expected} (\mu m)$ & $\Delta R_{measured} (\mu m)$ \\ \hline
$6.7 \pm 0.1$ & $62.0 \pm 0.9$ & $61.72 \pm 0.18$ \\ \hline
$6.9 \pm 0.1$ & $64.4 \pm 0.9$ & $64.01 \pm 0.23$ \\ \hline
$4.3 \pm 0.1$ & $39.7 \pm 0.9$ & $39.78 \pm 0.22$ \\ \hline
$4.4 \pm 0.1$ & $40.5 \pm 0.9$ & $40.02 \pm 0.21$ \\ \hline

\end{tabular}
\caption{Expected and measured distance changes versus temperature changes.}
\end{center}
\end{table}

\section{Conclusion}

In this paper,  high-precision absolute distance 
measurements were performed with frequency scanned interferometry
using dual-laser scanning and multi-distance-measurement techniques.
The dual-laser scanning technique is confirmed to cancel drift errors
effectively, and the multi-distance-measurement technique is used to
suppress the vibration and uncertainties from interference fringe
determination. Under realistic conditions, a precision of about 0.2 microns 
was achieved for an absolute distance of 0.41 meters.



This work is supported by the National Science Foundation and the Department of Energy of
the United States.








\begin{thebibliography}{99}

\bibitem{fsi05} Hai-Jun Yang, Jason Deibel, Sven Nyberg, Keith Riles, 
``High-precision Absolute Distance and Vibration Measurement using Frequency Scanned Interferometry'',
Applied Optics, Vol. 44, 3937-3944(2005). physics/0409110.

\bibitem{ox98} A.F. Fox-Murphy, D.F. Howell, R.B. Nickerson, A.R. Weidberg, 
``Frequency scanned interferometry(FSI): the basis of a survey system for ATLAS
using fast automated remote interferometry'',
Nucl. Inst. Meth. A383, 229-237(1996)

\bibitem{ox04} P.A. Coe, D.F. Howell, R.B. Nickerson,
   ``Frequency scanning interferometry in ATLAS: remote, multiple, simultaneous and precise distance measurements in a hostile environment'', Meas. Sci. Technol.15 (11): 2175-2187 (2004)

\bibitem{coe2001} P. A. Coe, 
``An Investigation of Frequency Scanning Interferometry for the alignment
of the ATLAS semiconductor tracker'', Doctoral Thesis, 
St. Peter's College, University of Oxford, Keble Road, Oxford, United Kingdom, 1-238(2001)

\bibitem{ilc} http://www.linearcollider.org/

\bibitem{orangebook} American Linear Collider Working Group(161 authors), 
``Linear Collider Physics, Resource Book for Snowmass 2001'', Prepared for the Department of Energy under contract
number DE-AC03-76SF00515 by Stanford Linear Collider Center, Stanford University, Stanford, California.
hep-ex/0106058, SLAC-R-570 299-423(2001) 

\bibitem{stone99} J.A. Stone, A. Stejskal, L. Howard, 
``Absolute interferometry with a 670-nm external cavity diode laser'',
Appl. Opt. Vol. 38, No. 28, 5981-5994(1999)

\bibitem{dai98} Dai Xiaoli and Seta Katuo, 
``High-accuracy absolute distance measurement by means of wavelength
scanning heterodyne interferometry'',
Meas. Sci. Technol.9, 1031-1035(1998)

\bibitem{barwood98} G.P. Barwood, P. Gill, W.R.C. Rowley, 
``High-accuracy length metrology using multiple-stage swept-frequency interferometry
with laser diodes'', Meas. Sci. Technol. 9, 1036-1041(1998)

\bibitem{bechstein98} K.H. Bechstein and W Fuchs, 
``Absolute interferometric distance measurements applying a variable synthetic wavelength'',
J. Opt. 29, 179-182(1998)

\bibitem{thiel95} J. Thiel, T. Pfeifer and M. Haetmann,
``Interferometric measurement of absolute distances of up to 40m'', Measurement 16, 1-6(1995)

\bibitem{kikuta86} H. Kikuta, R. Nagata, 
``Distance measurement by wavelength shift of laser diode light'', 
Appl. Opt. Vol. 25, 976-980(1986)


\end{thebibliography}
\end{document}